\documentclass[12pt]{iopart}

\usepackage{enumitem}
\usepackage{natbib}
\usepackage{amssymb}
\usepackage{graphicx}
\usepackage{color}
\usepackage{amsthm}
\usepackage{slashed}
\usepackage{soul}
\usepackage{amsbsy}

\begin{document}

\title[Locally Scale-Invariant Gravity]
{Locally Scale-Invariant Gravity}

\author{Meir Shimon}

\address{School of Physics and Astronomy, Tel Aviv University, Tel Aviv 69978, Israel}
\ead{meirs@tauex.tau.ac.il}
\vspace{10pt}

\begin{abstract}
We put forward the idea that in addition to diffeomorphism 
invariance of general relativity (GR) the gravitational 
interaction is invariant under arbitrary scale-deformations 
of the metric field. In addition, we assume that 
the scaling field has an internal symmetry. 
The global charges that are associated with this 
symmetry could potentially source the gravitational 
field. 
In the case that isotropic deformations are considered, 
the theory reduces to a Weyl-invariant (WI) version of 
GR. In the case that Minkowski spacetime is deformed 
the vierbein formalism is recovered, rendering GR a 
field theory on Minkowski spacetime. 
A few implications of a classical Weyl-invariant 
scalar-tensor (WIST) generalization of general 
relativity (GR) are considered. As an example, we 
recast the homogeneous and isotropic 
Friedmann-Robertson-Walker (FRW) spacetime in the WIST form 
with static space and monotonically evolving masses. 
\end{abstract}

\section{Introduction}
It is hard to conceive any physically realistic description 
of the Universe in the absence of either inertia or gravitation.
Inertia is thought to be a universal phenomenon; 
the Higgs field that is responsible for the 
generation of particle masses in the electroweak sector is 
ubiquitous. Likewise, gravitation is everywhere. It is normally 
identified with the omnipresence of spacetime, the substratum of 
physical reality, at least down to spacetime singularities, where 
the very concepts of space and time are believed to break down. 

Our current best theory of the (classical) gravitational interaction 
is general relativity (GR). The latter has successfully 
passed a wide range of tests in our solar system. 
GR is one of the simplest of all conceivable metric theories; 
it obeys the metricity condition, it is torsionless and 
it is described by a lagrangian linear in the Ricci scalar.
GR provides a diffeomorphism-invariant field theoretical framework 
that incorporates the principle of relativity applied to non-inertial 
observers and is realized on curved spacetimes. 
The curvature in the presence of a given 
energy-momentum distribution is governed 
by $G$, the universal coupling constant of gravitation that 
converts energy-momentum density units to length units. This 
specific coupling between matter and geometry was adopted in GR 
so as to guarantee consistency with Newton law of 
gravitation in the weak field limit, assuming that the 
locally inferred $G$ is universal.

Naturally, GR is expected to depart from the Newtonian limit 
in the strong gravity regime, i.e. in physical systems which are 
dominated by the gravitational interaction while other interactions 
play a secondary role, e.g. in black holes or on the largest 
cosmological scales. However, in contrast to expectations 
Newtonian gravity 
does not adequately describe gravitational phenomena on 
galactic scales and larger, e.g. at the outskirts of galaxies and 
galaxy clusters, even though gravity is weak on these scales, 
i.e. the dimensionless gravitational potential is as low as 
$\Phi\sim 10^{-4}$ or $10^{-3}$. 
A straightforward remedy for this 
problem (and others in cosmology) involves introducing some form of 
non-relativistic (NR) invisible matter, i.e. `dark matter' (DM). 
By far the leading 
candidate is some yet-to-be-determined beyond-the standard-model (SM) 
particle. According to another, less popular, possibility primordial 
black holes may account for the missing non-luminous matter.

As of yet, no DM particle candidate has been found in spite 
of decades of intensive search. In addition, the status of 
beyond-the-SM theories of particle physics -- our best hope for 
new DM particles -- is unclear with persistent null results coming from 
the large hadron collider (LHC), although certain anomalies 
have been reported recently with mild indications for new physics.
A recent high precision measurement of the Z boson mass 
at the Fermilab Tevatron collider, that significantly 
departs from SM expectations, calls the integrity of 
the SM into question.
In any case, these are either inconclusive as of yet or need independent 
corroborations by other experiments, and even if eventually ratified 
considerable amount of effort is still needed before the existence of 
a DM particle that adequately accounts for DM phenomena on Hubble down 
to galactic scales is compellingly established.

Another possibility is that particulate DM simply 
does not exist, and that on galactic scales or larger 
gravitation is not strictly described 
by GR, but rather by a modified version thereof. Indeed, the validity of GR 
coupled to purely ordinary matter (described by the SM of particle physics) 
has been only compellingly established within our own solar system. 
One possible extension of GR, invokes a higher symmetry -- Weyl 
invariance (WI), where $G$ is determined by a scalar field 
(while the other three fundamental interactions are assumed 
to have their standard forms, i.e. they are non-Weyl invariant). 
In GR this scalar field is replaced by default with a fixed $G$ 
throughout space and time. 
However, its {\it local} determination by Cavendish-type experiments does not 
necessarily imply its universality. In a WI version of GR, $G$ can vary 
in space and time, freely departing from its locally measured value, 
and simultaneously inducing scalar metric perturbations. 
From this perspective, there is {\it a priori} no preferred 
conformal frame, and whereas the choice $G=const.$ seems 
to work impressively well in our solar system it clearly 
fails on galactic scales and larger. 
It has been shown recently, that fractional variations of $G$ and active 
gravitational masses at the $O(10^{-3})$ or 
$O(10^{-4})$ level on galactic and 
galaxy cluster scales are sufficient for explaining away DM phenomena 
without particulate DM at the reasonable `cost' of endowing GR with WI [1]. 
Another application of a WI version of GR is a possible resolution 
of the `Hubble tension': the apparent statistically significant 
$\gtrsim 4\sigma$ tension between local and cosmological inference of 
the Hubble constant, $H_{0}$ [2]. 

On the largest cosmological scales observations require 
some form of a non-clustering vacuum-like DE, which is puzzling 
in a few respects. First, it is $\sim 120$ orders of magnitude 
suppressed relative to naive expectations based on our current 
understanding of particle physics. This might be hinting towards a 
gravitational/geometric origin for DE rather than material one.
Second, it is found empirically to be comparable in energy 
density terms to that of the NR matter at the present time 
although the two species greatly differ in their evolution 
history. In this sense, there seems to be a fine-tuning 
of the initial conditions that are associated with DE [3]. 

The present work focuses on the foundations of a certain 
generalization of GR that allows for a fully anisotropic 
locally scale-invariant generalization 
of GR by introducing a scale deformation (tensor) field 
$\Phi_{\mu\nu}$ living on an arbitrary 
spacetime. In the special case that the latter is Minkowski $\Phi_{\mu\nu}$ 
reduces to the well-known vierbein field. In the case of an 
isotropic scale-invariance on an arbitrary background 
the theory becomes a WI scalar-tensor (WIST) 
version of GR, hereafter referred to in abbreviation as `WIST'.
In addition to local scale-invariance we endow $\Phi_{\mu\nu}$ with an 
internal symmetry with which a global conserved charge is 
associated. The latter provides an additional source for 
spacetime curvature.

Although various variants of WIST have been 
contemplated over the past fifty years, e.g. [4-15], objections to 
this framework range between two extremes. On the one hand, it 
is often claimed that the WIST is 
`GR in a guise', that extending the symmetry of GR to allow WI does not make 
it genuinely so. In particular it is argued that there is 
classically no preferred conformal frame, so WIST is 
equivalent to GR. 
On the other hand, it is sometimes implied that dressing GR with 
WI brings about a ghost scalar field that inflicts `disastrous 
instabilities' on the theory, so much so as to disqualify the theory 
(while the same is not similarly claimed about the `equivalent' 
GR). 
We address these claims in the present work, and as a side we explicitly 
show that, not surprisingly, the action of the Friedmann-Robertson-Walker 
(FRW) spacetime, a legitimate solution of the Einstein equations, 
can be presented in a WIST form. 
By doing so, we illustrate with a specific example, with well-known 
general-relativistic description of spacetime, that there are no 
`fatal instabilities' associated with WIST that do not already exist 
in GR. Again, we stress that it is our assumption that WIST is entirely 
classical to the extent that GR is, and indeed the validity of GR has only 
been experimentally established on macroscopic down to $\mu m$, 
i.e. mesoscopic, scales, and not smaller.

The paper is organized as follows. The general anisotropic 
scale-invariant version of GR is considered in section 2, 
followed by its special isotropic case, the WIST theory, in section 3. 
The FRW example is discussed in section 4, followed by 
endowing the theory with global symmetries in section 5 and a 
summary in section 6. In Appendix A the transformation 
from a purely gravitational to non-gravitational description 
of the weak field approximation is described, and in Appendix B 
the case of U(2) global symmetry is briefly discussed.
Throughout, we adopt a mostly-positive signature for the 
spacetime metric $(-1,1,1,1)$, with the speed of light 
$c$ and reduced Planck constant $\hbar$ set to unity.

\section{Locally anisotropic scale-invariant theory of gravitation}

GR is described by the diffeomorphism-invariant Einstein-Hilbert (EH) action 
\begin{eqnarray}
\mathcal{I}_{EH}=\int\left(\frac{1}{16\pi G}(R-2\Lambda)
+\mathcal{L}_{M}\right)\sqrt{-g}d^{4}x, 
\end{eqnarray}  
where $R$ is the Ricci curvature scalar (obtained from the metric field 
$g_{\mu\nu}$ and its first and second derivatives in the standard way) 
and $\Lambda$ is the cosmological 
constant, respectively, $\mathcal{L}_{M}$ is the lagrangian density 
of matter, and $g$ is the metric determinant. The cosmological 
constant, $\Lambda$, can be absorbed in $\mathcal{L}_{M}$ as a species 
with vacuum-like properties. Diffeomorphism-invariance implies that equation (1) 
is invariant under the transformation 
$g_{\mu\nu}\rightarrow g_{\alpha\beta}\frac{\partial x^{\alpha}}
{\partial x^{\mu}}\frac{\partial x^{\beta}}{\partial x^{\nu}}$ 
(where Greek indices stand for spacetime coordinates and 
repeated indices are summed over). This symmetry of equation (1) 
under change of coordinate basis relies on the commutativity 
of derivatives. 

A natural generalization of diffeomorphism (that to the best of our 
knowledge has not been considered so far in the literature) 
could be the following replacement everywhere in equation (1)
\begin{eqnarray}
g_{\mu\nu}&&\rightarrow\bar{g}_{\mu\nu}\equiv \Phi_{\mu}^{\ \alpha}g_{\alpha\beta}\Phi_{\ \nu}^{\beta},\nonumber\\
\mathcal{L}_{M}&&\rightarrow\bar{\mathcal{L}}_{M}/|\Phi|,
\end{eqnarray} 
where $\Phi_{\mu}^{\ \alpha}(x)$ is an arbitrary spacetime-dependent 
real scale-deformation tensor of rank 2, 
and $|\Phi|\equiv\det(\Phi_{\mu}^{\ \nu})$.
This transformation represents a local {\it anisotropic} 
rescaling of the metric field and the lagrangian density 
(in which $\Lambda$ is absorbed), thereby 
providing us with a generalization of the 
`isotropic' local rescaling, known also as WI, that will be discussed 
below and in the next section.
In this framework gravitation is described by both a tensor 
field $\Phi_{\mu\nu}$ and the spacetime metric 
$g_{\mu\nu}$ on which $\Phi_{\mu\nu}$ lives.
We assume that $\Phi_{\mu\nu}$ is all-pervading 
much like the Higgs field is universal.

Naturally, the local (in general anisotropic) rescaling of the 
various fields is by physical dimension, e.g. the dimensions of 
$g_{\mu\nu}$, $g^{\mu\nu}$ and $\mathcal{L}_{M}$ are 
$length^{2}$, $length^{-2}$ and $length^{-4}$, respectively. 
The contravariant coordinates, $x^{\mu}$, are dimensionless, and 
the covariant coordinates $x_{\mu}$ are of dimension $length^{-2}$.
Specifically, the anisotropic local scale transformations for the metric and 
tensor fields, as well as the lagrangian density, are
\begin{eqnarray}
g_{\mu\nu}&\rightarrow&\Omega_{\mu}^{\ \alpha}g_{\alpha\beta}\Omega^{\beta}_{\ \nu}\nonumber\\ 
\Phi_{\mu}^{\ \alpha}&\rightarrow&\Phi_{\mu}^{\ \rho}(\Omega^{-1})_{\rho}^{\ \alpha}\nonumber\\
\Phi_{\ \mu}^{\alpha}&\rightarrow&(\Omega^{-1})_{\ \rho}^{\alpha}\Phi_{\ \mu}^{\rho}\nonumber\\
\mathcal{L}_{M}&\rightarrow &\mathcal{L}_{M}/|\Omega(x)|,
\end{eqnarray}
where $|\Omega(x)|$ is the determinant of an arbitrary spacetime-dependent 
matrix $\Omega_{\mu}^{\ \nu}$, 
leaves $\bar{g}_{\mu\nu}$ and $\bar{\mathcal{L}}_{M}/|\Phi|$ that are defined 
in equation (2), invariant. Therefore, if we replace 
$g_{\mu\nu}$ and $\mathcal{L}_{M}$ with $\bar{g}_{\mu\nu}$ 
and $\bar{\mathcal{L}}_{M}$, respectively, everywhere in equation (1) then 
the latter is invariant under equation (3) in addition to being 
diffeomorphism-invariant. 

In the special case that the scaling matrix is isotropic, 
$\Phi_{\mu}^{\ \nu}=\phi(x)\delta_{\mu}^{\ \nu}$, i.e. the 
metric scale deformation is isotropic, the proposed 
symmetry reduces to a WI version of GR, i.e. the aforementioned 
WIST theory, in which case the replacement 
$g_{\mu\nu}\rightarrow\tilde{g}_{\mu\nu}=g_{\mu\nu}\phi^{2}(x)$ 
and $\mathcal{L}_{M}\rightarrow\mathcal{L}_{M}/\phi^{4}$ 
is made in equation (1), where $\phi(x)$ is an arbitrary scalar function. 
This implies that 
the theory is invariant under the Weyl transformation
\begin{eqnarray}
g_{\mu\nu}&\rightarrow &\tilde{g}_{\mu\nu}=g_{\mu\nu}\Omega^{2}(x)\nonumber\\
\phi&\rightarrow &\tilde{\phi}=\phi\Omega^{-1}(x)\nonumber\\
\mathcal{L}_{M}&\rightarrow &\tilde{\mathcal{L}}_{M}
=\mathcal{L}_{M}\Omega^{-4}(x).
\end{eqnarray}
This isotropic case will be further explored in the next section.

In the special case 
$\eta_{\mu\nu}\rightarrow e_{\mu}^{\ \alpha}\eta_{\alpha\beta} 
e_{\ \nu}^{\beta}=g_{\mu\nu}(x)$, 
where $\eta_{\mu\nu}$ is the Minkowski metric, the tensor 
$\Phi_{\mu}^{\ \nu}=e_{\mu}^{\ \nu}$ is the vierbein field. 
Tensor-mode perturbations, i.e. gravitational waves, 
could therefore describe massless tensor perturbations 
in the $\Phi_{\mu}^{\ \nu}$ field, 
i.e. in `G' and in active gravitational masses, rather than in spacetime itself.
When spacetime is fixed to be Minkowski all the gravitational 
degrees of freedom that are normally represented by the various 
components of the spacetime metric are now accounted for by the tensor 
field, which in this case is merely the vierbein field, $e_{\mu}^{\nu}$. 
In this choice the metric field itself, that describes Minkowski spacetime, 
is non-dynamical, space and time are infinite and and all the 
singularities are in $\Phi_{\mu}^{\ \nu}$, its derivatives, and 
combinations thereof that replace, e.g. the standard spacetime 
curvature. This differs from the standard interpretation of GR 
where distortions and singularities are understood to be 
in spacetime itself, where e.g. the Big Bang singularity 
is normally understood as the beginning of time, and perhaps 
even of spacetime. To illustrate the description of gravitational 
phenomena within the WIST theory on Minkowski spacetime we examine 
the transformation of the geodesic equation for a point test particle 
in the `weak field' approximation in Appendix A. 

Setting $G=3/(8\pi)$ and replacing $\mathcal{L}_{M}$ with $V$ 
(the latter contains no derivatives of the various fields), and 
$g_{\mu\nu}$ with $\bar{g}_{\mu\nu}$ (see equation 2) everywhere in equation (1), 
we obtain
\begin{eqnarray}
\bar{\mathcal{I}}=\int\left(\frac{1}{6}(\bar{R}-2\Lambda)
+V\right)\sqrt{-\bar{g}}d^{4}x. 
\end{eqnarray} 
As will be detailed below we assume that $V$, the source of spacetime curvature, 
is a potential in $\Phi\equiv\sqrt{\Phi_{\alpha}^{\beta}\Phi_{\beta}^{\alpha}}$. In particular, we make the assumption 
that it is an analytic polynomial in $\Phi$. 
Consider the transformations described by equation (3). 
To first order, an infinitesimal transformation results in
\begin{eqnarray}
\bar{g}_{\mu\nu}=\Omega_{\mu}^{\ \alpha}g_{\alpha\beta}
\Omega_{\ \nu}^{\beta}\approx g_{\mu\nu}+\delta\Omega_{\mu}^{\ \alpha}g_{\alpha\nu}+g_{\mu\alpha}\delta\Omega_{\ \nu}^{\alpha},  
\end{eqnarray}
where we used $\Omega_{\mu}^{\ \nu}\approx\delta_{\mu}^{\ \nu}+\delta\Omega_{\mu}^{\ \nu}$. Similarly,
\begin{eqnarray}
\bar{\Phi}_{\mu}^{\ \nu}=(\Omega^{-1})_{\mu}^{\ \alpha}
\Phi_{\alpha}^{\ \nu}\approx (\delta_{\mu}^{\alpha}
-\delta\Omega_{\mu}^{\ \alpha})\Phi_{\alpha}^{\ \nu}, 
\end{eqnarray}
from which it follows that 
\begin{eqnarray}
\delta\Phi_{\mu}^{\ \nu}\approx 
-\delta\Omega_{\mu}^{\ \alpha}\Phi_{\alpha}^{\ \nu}. 
\end{eqnarray}
Employing the relation between variations of inverse matrices, 
$\delta g^{\mu\nu}=-g^{\mu\alpha}\delta g_{\alpha\beta}g^{\beta\nu}$ 
and using equation (6) we obtain
\begin{eqnarray}
\delta g^{\mu\nu}\approx -g^{\mu\alpha}\delta\Omega_{\alpha}^{\ \nu}-\delta\Omega_{\ \alpha}^{\mu}g^{\alpha\nu}. 
\end{eqnarray}
In our convention that $x^{\mu}$ are dimensionless (see discussion below equation 2), 
$\int V\sqrt{-g}$ is {\it dimensionless} 
and so $\delta (V\sqrt{-g})=\frac{(\delta V\sqrt{-g})}{\delta g^{\mu\nu}}
\delta g^{\mu\nu}
+\frac{(\delta V\sqrt{-g})}{\delta\Phi_{\mu}^{\ \nu}}\delta\Phi_{\mu}^{\ \nu}
+\frac{(\delta V\sqrt{-g})}{\delta\Phi_{\ \mu}^{\nu}}\delta\Phi_{\ \mu}^{\nu}=0$
under the transformations Eqs.(8) and (9). Requiring invariance under arbitrary 
$\delta\Omega_{\mu}^{\ \nu}$ we obtain
\begin{eqnarray}
S_{\mu}^{\ \nu}=\Phi_{\mu}^{\ \alpha}\frac{\delta V}{\delta\Phi^{\alpha}_{\ \nu}}
+\frac{\delta V}{\delta\Phi^{\ \alpha}_{\nu}}\Phi_{\ \mu}^{\alpha}
\end{eqnarray}
where $S_{\mu\nu}\equiv -\frac{2}{\sqrt{-g}}
\frac{(\delta V\sqrt{-g})}{\delta g^{\mu\nu}}$.
This provides a strong constraint on the potential if anisotropic local 
scale invariance, Eqs. (1)-(3), is to be a symmetry of the gravitational 
interaction. We will rederive the analog of this result 
in the isotropic scaling case and delve into the implications 
in sections 3 and 4.

We emphasize that only gravitation is 
endowed with WI in the proposed framework. All other 
fundamental interactions, namely the strong and electroweak 
interactions are described by the SM of particle 
physics, with no modifications. 
Since any realistic physical system is governed by 
both gravitational and non-gravitational interactions, 
and since the fundamental interactions are 
minimally-coupled to gravitation, it then follows that this 
particular framework is clearly not merely a field re-definition 
of GR and the SM of particle physics. In any case, introducing the global 
charge that is generally associated with $\Phi_{\mu\nu}$ 
due to internal symmetry as discussed in section 5 and in Appendix B 
renders the theory described here inequivalent to the SM even in purely 
gravitational systems such as late time cosmological scales.

Much like the nonvanishing 
and universality of VEV of the Higgs field guarantees that inertia 
phenomena are ubiquitous so it is the case with spacetime that 
never ceases to exist as is reflected by the fact that the metric 
field can always be so chosen to be Minkowski in the most general 
anisotropic locally scale-invariant gravitation. Since the SM of particle 
physics is minimally coupled to gravitation then the tensor field 
$\Phi_{\mu\nu}$ locally couples to the SM fields instead of them 
being living on a curved spacetime. For example, 
$\mathcal{I}_{kin}=\int g^{\mu\nu}(D_{\mu}H)^{\dagger}D_{\nu}H\sqrt{-g}d^{4}x$, 
the kinetic term of the Higgs field $H$ 
(where $D_{\mu}$ is the gauge-covariant derivative), transforms to
$\mathcal{I}_{kin}=\int (D_{\mu}H)^{\dagger}\bar{g}^{\mu\nu}D_{\nu}H
\sqrt{-\bar{g}}d^{4}x$, which in the Minkowski background simplifies to 
$\mathcal{I}_{kin}=\int (D_{\mu}H)^{\dagger}\Phi^{\mu}_{\ \sigma}
\Phi^{\sigma\nu}D_{\mu}H|\Phi|d^{4}x$. The tensor field $\Phi^{\mu\nu}$ 
then directly couples to $H$, 
(where $\Phi_{\mu}^{\ \nu}$ is actually the vierbein field in the case 
of Minkowski background) 
thereby constituting a field theory on Minkowski spacetime 
(spacetime indices are raised and lowered by means of the Minkowski metric), 
as opposed to the standard interpretation according to which $H$ interacts 
with the geometry of spacetime.

\section{WIST theory}
In this section we focus on the isotropic case, i.e. 
$\Phi_{\mu}^{\nu}=\delta_{\mu}^{\nu}\phi(x)$, where 
$\phi(x)$ is a scale deformation (scalar) field. 
Although it was assumed in equation (2) that $\Phi_{\mu}^{\nu}$ is real 
(in order for the new metric field, $\bar{g}_{\mu\nu}$, 
to be real) it is clear that at least in the special 
case considered in this section, i.e. 
$\Phi_{\mu}^{\nu}=\delta_{\mu}^{\nu}\phi(x)$,
promoting $\phi(x)$ to a complex scalar field renders 
$\bar{g}_{\mu\nu}$ real. Since the scalar field is assumed 
to be complex, it has a U(1) internal symmetry.
The more general case of an arbitrary internal symmetry is 
briefly discussed in the section 5. Gravitation is 
then fundamentally endowed with WI, i.e. an {\it isotropic} locally 
scale-invariance, within a WIST theory, and standard GR is recovered 
in a particular conformal frame, where $\phi=constant$.

Rather than using the more general results from the previous section 
we derive them directly in the isotropic case.
The WIST theory readily follows from locally rescaling the 
spacetime metric $g_{\mu\nu}\rightarrow \phi\phi^{*}g_{\mu\nu}$ 
everywhere in equation (1). The resulting theory is then described by 
the following action
\begin{eqnarray}
\mathcal{I}_{WIST}&=&\int\left(\frac{1}{6}|\phi|^{2}R
-\phi^{*}\Box\phi+V(|\phi|,\{\Psi\})\right)
\sqrt{-g}d^{4}x\nonumber\\
&=&\int\left(\frac{1}{6}|\phi|^{2}R
+\phi_{\mu}\phi^{*\mu}+V(|\phi|,\{\Psi\})\right)\sqrt{-g}d^{4}x,
\end{eqnarray}
where $\phi_{\mu}\equiv\frac{\partial\phi}{\partial x^{\mu}}$, 
$V$ that replaces $\mathcal{L}_{M}$ 
is now allowed to explicitly depend on $|\phi|$ 
but not on its derivatives, 
and the second equality follows from integration by parts 
of the kinetic term associated with the scalar field.
All other fields (including $g_{\mu\nu}$) 
are collectively denoted by $\{\Psi\}$.
We emphasize that in the new formulation, equation (11), 
gravitation is sourced by $V$ rather than by `matter', 
$\mathcal{L}_{M}$. 
In general, GR is consistent with observations 
within the solar system and so on these scales 
$V$ can be replaced by $\mathcal{L}_{M}$, i.e. `ordinary' matter. 
However, on larger scales, where both DM and DE are 
required by observations we favor the form equation (11) 
which does not imply the existence of DM particles. 
In addition, in equation (11) 
geometry and matter are not presented as two different 
entities. Rather, the existence of $\phi$ in both the 
source, $V$, and the kinetic term, reflects the viewpoint 
that they are actually intertwined. They are 
`unified' in the sense that WI of equation (11), a symmetry 
that is not recognized by GR, depends on both the scalar and 
metric fields.

With $V$ being explicitly $\phi$-dependent, equation (11)
is now a scalar-tensor theory of the Bergmann-Wagoner type [16], [17].
It was first obtained by Deser [4] and 
later in [10] with $\mathcal{L}_{M}$ 
[referred to as V in equation (11)] that does not necessarily 
depend on $\phi$. In [4] $\phi$ was assumed to be real.
A similar procedure to the replacement 
$g_{\mu\nu}\rightarrow \phi\phi^{*}g_{\mu\nu}$ that 
transforms equation (1) to equation (10) was employed 
by Chamseddine and Mukhanov in their Mimetic Gravity 
[18] but in the latter case the scalar field was 
an irotational velocity potential $\varphi$, i.e. 
$g_{\mu\nu}\rightarrow \varphi_{,\mu}\varphi^{,\mu}g_{\mu\nu}$.

Formally, the kinetic term associated with the scalar field that 
appears in equation (11), 
$\mathcal{L}_{\phi}\equiv\phi_{,\mu}\phi^{*,\mu}$ or 
$\mathcal{L}_{\phi}\equiv -\phi^{*}\Box\phi$, 
can be viewed as a new source of the gravitational interaction. This 
term is completely ignored in GR where $\phi$ is set to a constant value, 
$\phi_{0}\equiv\sqrt{\frac{3}{8\pi G}}$. 
Invariance of equation (11) implies that $V\rightarrow|\phi|^{-4}V$ under $g_{\mu\nu}\rightarrow |\phi|^{2}g_{\mu\nu}$. 
This simple derivation also underscores the origin of the `ghost' 
scalar field whose kinetic term, 
$\mathcal{L}_{\phi}\equiv\phi_{,\mu}\phi^{*,\mu}$, 
appears with the `wrong' sign in equation (11); 
it simply stems from locally stretching/squeezing the 
spacetime metric, or equivalently stretching/squeezing the yardstick 
with which distances are measured, in our case it is the Planck 
length $l_{p}\propto\sqrt{G}$ in vacuum, or $\propto (G\rho)^{-1/2}$ 
where $\rho$ is the energy density in non-vacuum configurations. 
This observation is crucial for evaluation of 
the significance of the ghost field. We see that $\phi$ 
is tightly related to $g_{\mu\nu}$ (merely a local stretch 
of the spacetime metric), and as $g_{\mu\nu}$ is treated 
{\it classically} so should be $\phi$ and $V$; the ghostly nature of $\phi$ 
derives solely from the specific form in which $g_{\mu\nu}$ and 
its derivatives appear in the EH action. Below, we explicitly show 
(although it should be clearly evident already from the very local 
scaling procedure $g_{\mu\nu}\rightarrow\phi\phi^{*}g_{\mu\nu}$ that 
the spacetime dependence of $\phi$ is arbitrary)
that the {\it classical} field $\phi$ is not determined by any 
{\it dynamical} equation (indeed not unexpected in a WI theory). 
In particular, it is not driven {\it dynamically} to unstable 
field configurations (which is the generic problem with 
ghost fields in non-WI theories). The latter are considered 
disastrous in the quantum regime because the Hamiltonian is 
unbounded from below.

Since equation (11) was obtained from equation (1) by 
affecting the transformation 
$g_{\mu\nu}\rightarrow\phi\phi^{*}g_{\mu\nu}$, 
invariance of the former is guaranteed insofar 
$|\phi|^{2}g_{\mu\nu}$ is invariant, i.e. 
\begin{eqnarray}
\phi&\rightarrow&\phi/\Omega\nonumber\\
g_{\mu\nu}&\rightarrow&\Omega^{2}g_{\mu\nu},
\end{eqnarray}
for an {\it arbitrary} $\Omega(x)>0$, i.e. equation (4).
Equation (12) are the Weyl transformations of metric 
and scalar fields, respectively.
It is thanks to this freedom 
that $\phi=\phi_{0}/\Omega(x)$ is arbitrary, and so whereas it is 
spacetime-dependent it is {\it not} dynamical in the sense that 
it is {\it not} determined by a dynamical differential equation which 
has an attractor solution, that, e.g. drives the kinetic term, 
$\mathcal{L}_{\phi}\equiv\phi_{,\mu}\phi^{*,\mu}$, 
to arbitrary negative values, as is the case with generic 
ghost fields. Actually, we will see in section 4 
that exactly the opposite takes place in our expanding Universe when 
the action of the FRW spacetime is 
recast in the form of equation (11) instead of equation (1); 
the negative kinetic term evolves from negative infinity at the Big Bang 
towards vanishingly negative values at the remote future assuming a 
monotonic expansion history, compatible with the concordance 
cosmological model.

In fact, the presence of the kinetic term in 
$\mathcal{L}_{\phi}\equiv\phi_{,\mu}\phi^{*,\mu}$ is expected once 
$G$ is promoted to a field. More specifically, 
since the curvature scalar depends on derivatives of 
the metric field it transforms inhomogeneously 
under equation (12)  
\begin{eqnarray}
R\rightarrow\Omega^{-2}\left(R-6\frac{\Box\Omega}{\Omega}\right), 
\end{eqnarray}
where $\Box$ is the d'Alambertian. Invariance of equation (11) 
under equation (12) then requires the presence of the kinetic term 
$\mathcal{L}_{\phi}\equiv\phi_{,\mu}\phi^{*,\mu}$ that transforms inhomogeneously 
and guarantees the mutual cancellation of derivatives of $\Omega$ 
(provided that the appropriate integration by parts has been 
carried out). The prefactor $\frac{1}{6}$ in front of the curvature 
coupling term in equation (11) guarantees that 
the inhomogeneous term on the right hand side of equation (13) 
is compensated by a similar inhomogeneous term from the transformation 
of the kinetic term, $\phi^{*}_{\mu}\phi^{\mu}$. In other words, 
under the transformation equation (12) the combination of the 
first two terms in equation (11) transforms as, 
$\frac{1}{6}|\phi|^{2}R-\phi\Box\phi^{*}\rightarrow 
\Omega^{-4}(\frac{1}{6}|\phi|^{2}R-\phi\Box\phi^{*})$, i.e. it 
has the appropriate well-defined conformal weight of 
$length^{-4}$, such that its product with the volume element in 
equation (11) is WI.

In the special case that $\phi$ is a real field and 
V is independent of $\phi$ equation (11) can be cast 
in the form of a Brans-Dicke (BD) theory [19] 
$\mathcal{I}_{BD}=\int\Big(\Phi R
-\omega_{BD}\frac{\Phi_{\mu}\Phi^{\mu}}{\Phi}+\mathcal{L}_{M}\Big)
\sqrt{-g}d^{4}x$ with $\Phi$ (not to be confused with the 
norm of $\Phi_{\mu\nu}$ that is defined just below equation 5) and 
$\omega_{BD}$ being the BD scalar field 
and dimensionless BD parameter, respectively, 
where $\Phi\equiv\phi^{2}/6$ and $\omega_{BD}\equiv -3/2$. 
Whereas a tight lower limit, $\omega_{BD}>40000$, has been reported in 
[20], it should be stressed that it was obtained from satellite 
observations directly exploring the impact of varying $G$ within our 
own solar system, and not beyond. In addition, in the (generally 
non-WI) BD framework the source term $\mathcal{L}_{M}$ does not 
in general depend on $\phi$ while in WIST the potential $V$ does.

It is constructive at this point to further elucidate 
the issue of stability 
of Eq. (11) unlike what is sometimes claimed or implied. 
While it is true that the kinetic term appears 
in equation (11) with the 
`wrong' sign relative to $V(|\phi|)$ and $\phi$ is formally a `ghost' 
field it is merely a gauge artifact that can be seen when 
the latter is set to a constant in the 
unitary gauge [21]. The `fatal instability' usually ascribed 
to quantum ghost fields stems from the 
fact that in their presence there is either no stable 
vacuum state, or non-conservation of probability takes place, 
e.g. [22]. However, GR is classical, and so is WIST.
As was already mentioned above, 
in the special case of WIST, it is exactly WI that 
prevents classical instability from taking 
place because there is essentially no equation of motion for 
$\phi$ to {\it dynamically} drive it towards instability. 
This is not surprising since at least classically, 
two conformally-related theories are 
equivalent, e.g. [23], [24]. Actually, it has been 
argued that WI is a sham/fake symmetry of GR [25], [26], i.e. 
that equation (11) is equivalent to GR. On face value, the conclusion 
that the scalar field is a `spurion' [25] implies that its 
formal ghostly nature is of no practical significance. As 
mentioned above, adding a global symmetry to $\Phi_{\mu\nu}$ 
renders both the anisotropic and isotropic (equation 11) locally 
scale invariant theories inequivalent to GR.
We also re-iterate in this context that in the framework 
proposed here gravitation is WI, whereas the SM of particle 
physics is not, and since in general 
the other three interactions are non-negligible in typical physical 
systems, then the SM of particle physics coupled to WIST 
is {\it not} equivalent to the SM of particle physics coupled to GR. 
This was a key observation that was exploited in [1] in explaining 
away particulate DM in galaxies and clusters of galaxies. Had the SM 
of particle physics been WI as well then the proposed remedy for the DM 
problem would gauge away.
 
Ultimately, the `appropriate' conformal frame to be used in the description 
of a physical system should be determined by observations, much like the 
`appropriate' coordinate system employed in the description of, e.g., 
a black hole in GR is determined by observations, e.g. [24]. For example, a 
spherically symmetric static black hole as seen from the perspective 
of a static observer would be most naturally described by a Schwarzschild 
metric, while a freely falling observer towards the same black hole will 
likely employ a different, time-dependent, metric in describing the 
same black hole. The diffeomorphism symmetry of GR allows observers 
to describe the system in the most symmetric fashion, which clearly 
depends on the ({\it a priori} arbitrary) observers states. 
In the same fashion, at the classical level, theory cannot 
pre-select the `appropriate' units to be used in the description of a 
physical system. This freedom is clearly not a property of GR, but is 
certainly a desirable tenet of its WI generalization described 
by equation (11) or the more general, locally anisotropic 
scale-invariant, theory described in the previous section.

In the quantum regime 
the situation generally changes as quantization and Weyl 
transformations do not in 
general commute, e.g. [23], [24], and in fact ghosts are known to 
violate unitarity, e.g. [27]. However, quantization of 
GR has never been demonstrated to be feasible (at least this 
is the consensus), nor are there available measurements indicating 
that quantum gravitational effects are in play so as to necessitate 
its quantization.
Actually, the standard theoretical arguments in favor of its quantization 
might be strong, yet inconclusive (e.g. [28] and references within), 
and at the very best GR could be considered an effective low energy 
limit of a more fundamental quantum theory of gravity, 
if it should indeed be ultimately quantized.
Therefore, in the same fashion that the non-renormalizability 
of GR does not disqualify it from being the backbone of the standard 
cosmological model what might be a ghost `problem' of equation (11) 
at the quantum level is not an issue for physical 
systems that are adequately described by GR, e.g. [29]. 
Therefore, the inequivalence between conformal frames at 
the quantum level need not concern us in the present work. 

It has been claimed in [27] that fixing the scalar 
field [and thereby rendering  equation (11) equivalent to GR] 
in order to go around the `ghost problem' in practice `robbs' 
equation (11) from its claimed WI. Again, while this 
argument might be valid 
at the quantum level it definitely does not apply classically. 
Moreover, and as discussed below, the set of fields 
$\phi$ and $g_{\mu\nu}$ 
that appears in equation (11) is under-determined and $\phi$ could be chosen to 
be any arbitrary function, not necessarily of a fixed value, thereby 
still `manifesting' WI, 
while at the same time avoiding driving the field configuration 
away from its equilibrium state. 
In summary, the `wrong' sign of the scalar field in a WI theory is not an issue 
at the classical regime, which is the only domain at which 
equation (11), exactly as GR, can be trusted.

It is constructive at this point to explore a slightly more general case of 
a {\it dynamical} scalar field which is non-minimally coupled 
to spacetime curvature
\begin{eqnarray}
\mathcal{I}_{ST}=\int\Big(\xi|\phi|^{2}R
+\phi^{*}_{\mu}\phi^{\mu}+V(|\phi|)\Big)
\sqrt{-g}d^{4}x, 
\end{eqnarray}
where $\xi$ is an arbitrary dimensionless coupling parameter, 
thereby generalizing equation (11), i.e the theory is in 
general not WI, unless $\xi=\frac{1}{6}$.

The field equations derived from variation of equation (14) 
with respect to the metric and scalar field are, respectively
\begin{eqnarray}
2\xi|\phi|^{2}G_{\mu\nu}&=&S_{M,\mu\nu}+\Theta_{\mu\nu},
\end{eqnarray}
and
\begin{eqnarray}
\xi\phi R-\Box\phi+\frac{\partial V}{\partial\phi}&=&0,
\end{eqnarray}
where
\begin{eqnarray}
&&3\Theta_{\mu\nu}\equiv 6\xi\phi^{*}\phi_{;\mu;\nu}
+(6\xi-3)\phi^{*}_{\mu}\phi_{\nu}\nonumber\\
&-&6\xi g_{\mu\nu}\left[\phi^{*}\Box\phi
-\left(\frac{1}{4\xi}-1\right)
\phi^{*}_{\alpha}\phi^{\alpha}\right]+c.c.
\end{eqnarray}
Here and throughout $S_{M,\mu\nu}\equiv -\frac{2}{\sqrt{-g}}\frac{\delta(\sqrt{-g}V)}
{\delta g^{\mu\nu}}$ is the analog of the energy-momentum tensor.
Equation (15) is a generalization of Einstein equations 
with $2\xi|\phi|^{2}$ replacing 
$1/(16\pi G)$, and $\Theta_{\mu\nu}$ is an effective 
contribution to the energy-momentum tensor essentially 
due to gradients of G and active gravitational masses, 
the latter appear in the source term, $\mathcal{L}_{M}$.
Multiplying equation (16) 
by $\phi^{*}$, adding the 
result to its complex conjugate and to the trace of 
equation (15) results in
\begin{eqnarray}
\phi\frac{\partial V}{\partial\phi}+
\phi^{*}\frac{\partial V}{\partial\phi^{*}}=
S_{M}+(1-6\xi)(\phi^{*}\Box\phi+\phi^{*}_{\alpha}\phi^{\alpha}+c.c).
\end{eqnarray}
Together, equation (15) and equation (18) 
provide a system of dynamical equations 
for the ten metric components and the complex scalar field.
In the BD form ($\phi$ is real and $\mathcal{L}_{M}$, that is here 
replaced by $V$, is independent of $\phi$) equation (18) 
reduces to the well-known form $(3+2\omega_{BD})\Box\Phi=T_{M}$.
It follows from equation (18) that 
for any $\xi\neq\frac{1}{6}$ 
(equivalently $\omega_{BD}\neq -3/2$) 
the scalar field is {\it dynamically} determined by 
the matter distribution via a {\it differential equation}, 
which thereby embodies the Machianity of BD theory of gravity.

In contrast, in the special case 
$\xi=\frac{1}{6}$ (equivalently $\omega_{BD}= -3/2$ [4-15]) 
Equation (14) is identical to (11), and equation (18) 
reduces to a constraint 
that only determines the dependence of $V$ on 
the scalar field (as in the more general equation 10). 
In this special case, $\phi$ is indeed non-dynamical; 
{\it it can be set to any desired function}, 
irrespective of the matter distribution, and Machianity is lost. 
Although $\phi$ still appears in equation (15) we 
now have only ten differential equations for ten 
metric components and a scalar field -- this under-determined 
system simply reflects the freedom to locally rescale the fields 
in the special case $\xi=\frac{1}{6}$, merely a manifestation of 
WI, equation (12), [4-15]. 
The latter implies invariance of equation (11) 
under local-rescaling of fields such that 
$V\rightarrow|\phi|^{-4}V$. 
This requires in particular that, e.g.,
\begin{eqnarray}
\psi&\rightarrow&\Omega^{-\frac{3}{2}}\psi\nonumber\\
A_{\mu}&\rightarrow&A_{\mu},\nonumber\\
\end{eqnarray}
where $\psi$ and $A_{\mu}$ are Dirac and vector fields, 
respectively, in addition to the transformation of the scalar 
and metric fields described in equation (12) -- each 
field is rescaled by its mass/length dimension. 
In the special case $\xi=\frac{1}{6}$, 
equation (18) could be derived solely from the requirement 
that $V\sqrt{-g}$ is WI, as was done in the previous 
section in the more general case, in equation (10).

Any mass terms that appear in $V$ are by 
definition active gravitational masses, 
which need not be equivalent to inertial 
or passive gravitational masses. 
While the three types of mass 
are not necessarily equivalent [30], 
the notion of a passive gravitational mass in 
any theory of gravity that satisfies the equivalence principle 
is vacuous. If the ratio of the latter two mass types is a universal 
constant then the equivalence principle is satisfied -- an assumption 
that we indeed make in the present work. 
Again, we emphasize that the other three fundamental interactions are described by 
the lagrangian of the SM of particle physics, $\mathcal{L}_{SM}$, 
where all masses are inertial, whether generated via the Higgs mechanism in 
the electroweak sector or via the explicitly broken chiral symmetry in QCD.
 
Since $\phi(x)=\Omega^{-1}(x)\phi_{0}$, where $\phi_{0}$ is the fixed 
GR value, then for any given such a choice $\Theta_{\mu\nu}$  
can be calculated according to equation (17) 
and used in equation (15) to solve for the 
corresponding $g_{\mu\nu}$. In practice, however, it is easier to start from any 
known solution of the Einstein equations, $\phi_{0}$ 
and $g_{\mu\nu}$, 
and transform it to $\Omega^{-1}(x)\phi_{0}$ 
and $\Omega^{2}(x)g_{\mu\nu}$ 
with arbitrary $\Omega(x)$ by virtue of WI. 
We emphasize, once again, that this latter procedure is 
only possible assuming (as is virtually always assumed) 
that there are no global charges associated with 
the field as is discussed in section 5.

In the theory described by equation (11), 
the scalar field $\phi$ determines not only $G$ but also 
active gravitational masses.
Only pure radiation $S_{rad}=0$ is consistent with the theory 
described by equation (14) with $\xi=\frac{1}{6}$ unless $V$ 
explicitly depends on $|\phi|$.
In the case of perfect fluid $S=V$ on shell [31], 
and with EOS $w_{M}$ the trace is 
$S=-\rho_{M}(1-3w_{M})$. It then follows from 
equation (18) that $V\propto|\phi|^{1-3w_{M}}$ 
in case that $\xi=\frac{1}{6}$, i.e it is linear and 
quartic in $|\phi|$ in cases of NR- and vacuum-like terms, respectively, 
and is independent of $|\phi|$ in case of pure radiation. 
Linearity of $V$ in $|\phi|$ in the case of vanishing EOS, 
$w_{M}=0$, suggests that active gravitational masses are regulated 
by $|\phi|$. Not only that the same $\phi$ determines both Planck 
mass and active gravitational masses is a necessary condition 
for the consistency of non-radiation sources with this WI theory, 
it is also a conceptually natural ``conclusion'' as the 
concept of active gravitational mass is meaningless unless 
it couples to G, and in this sense it seems natural that 
both quantities are determined by the same scalar field. 
This clearly does not have necessarily to be the case in 
general (as in e.g, BD theory), but it is a nice 
merit of the WI model described by equation (11).

We emphasize that in the original BD proposal 
[19] the matter lagrangian, $\mathcal{L}_{M}$, does not 
explicitly depend on the scalar field. Notably, Brans \& Dicke 
required $\omega_{BD}>0$ to guarantee the positivity of the Hamiltonian in their 
original proposal [19], a fact that was emphasized and reinterpreted in [32].
The instability of BD theories with $\omega_{BD}<0$ was further 
emphasized in [33], 
but as we argued above, the non-positive kinetic term of the scalar field is 
not an issue in the {\it special case} $\omega_{BD}=-3/2$ and 
$V=V(|\phi|)$ due to WI.
The observational lower limit $\omega_{BD}\geq 40000$ reported 
in [20] implies that the BD field $\Phi$ is very nearly a constant, 
essentially reducing the theory to GR. But, again, this conclusion 
applies to the BD theory, where $\mathcal{L}_{M}$ is independent of 
the scalar field, and in addition (perhaps not less important and relevant 
to astrophysics and cosmology) this tight limit has been obtained 
in our solar system and clearly does not readily apply on, e.g., 
galactic scales.

\section{The FRW spacetime In WIST representation}
In the present section we cast the FRW action in the 
form of equation (11) in a particular conformal frame where 
spacetime is static. The conventional initial curvature 
singularity at the Big Bang is replaced in this alternative 
description by the vanishing of $\phi$, i.e. of Planck and 
active gravitational masses, at $t=0$. The redshifting Universe is then 
a reflection not of space expansion, but rather of temporal 
evolution of masses, i.e. the redshifting Universe is 
a manifestation of shrinking Planck length and Compton 
wavelengths. The latter is a result of the fact that the 
time coordinate is now the conformal time, $\eta$, 
rather than the cosmic time, $t$, as will be discussed below.

The FRW action
\begin{eqnarray}
\mathcal{I}_{FRW}&=&\int[-a'^{2}+Ka^{2}
-\Lambda a^{4}/3+a^{4}\mathcal{L}_{M}(a)]\sqrt{-\gamma}d^{4}x, 
\end{eqnarray}
is obtained from equation (1) in units where $3/(8\pi G)\equiv 1$ 
and assuming that the metric is 
$g_{\mu\nu}=a^{2}\gamma_{\mu\nu}$, where the two metrics 
$g_{\mu\nu}$ and $\gamma_{\mu\nu}$ are conformally related,
and $\gamma_{\mu\nu}\equiv 
diag(-1,\frac{1}{1-Kr^{2}},r^{2},r^{2}\sin^{2}\theta)$ 
is the metric in the comoving frame, 
$\gamma\equiv det(\gamma_{\mu\nu})$, 
and after the term proportional to the corresponding curvature 
scalar $a^{4}R=6a^{4}(a''/a+K)/a^{2}$ is 
integrated by parts. Here, $K$ is the spatial curvature parameter, 
$a(\eta)$ is the purely time-dependent scale factor, 
a prime denotes derivatives with respect to $\eta$, 
where $d\eta\equiv dt/a$ and $d^{4}x=d\eta dr d\theta d\phi$.
 
Since null geodesics are blind to conformal metric transformations we 
expect light to still be redshifted in the WIST description with the 
spatially-static metric $\gamma_{\mu\nu}$ now replacing the standard 
expanding space metric $g_{\mu\nu}$. This is explained by the fact 
that $-d\eta^{2}=-\frac{dt^{2}}{a^{2}}$, and so the the lapse function, 
$\gamma_{00}$, is $a^{-2}$ if the metric $\gamma_{\mu\nu}$ is presented in cosmic 
time, $t$, coordinates, rather than conformal $\eta$. Incoming photon wavelengths 
are not stretched by space expansion in this description (space is static) 
but rather by the temporaly-evolving gravitational potential, 
$\gamma_{00}(t)$, which is $\propto a^{-2}(\eta)$, which as 
we see below is $\propto \phi^{-2}(\eta)$. In other words, 
cosmological redshift is (indirectly) due to the growing 
(decreasing) Planck mass (length). A more direct way to explain redshift is 
by the fact that in the comoving frame, described by $\gamma_{\mu\nu}$, 
inertial masses $M_{0}$ are transformed to $M=a(\eta)M_{0}$, i.e. 
the Rydberg constant monotonically increases with time.

We thus see that the ratio between either the Planck or active 
gravitational mass to the inertial mass is time-independent 
on cosmological scales where the distribution of matter is assumed 
to be homogeneous and isotropic and Eq. (20) applies. 
However, we stress that this is not the case in general 
spacetimes and matter distributions 
since the inertial mass is independent of the scalar field. 
It is only due to the specific symmetries of the cosmological 
model making it amenable to a particular combination of Weyl 
(e.g. $g_{\mu\nu}\rightarrow\gamma_{\mu\nu}=g_{\mu\nu}/a^{2}$) 
and coordinate (i.e. $dt\rightarrow d\eta=dt/a$) transformations 
that, e.g. the ratio of inertial-to-Planck mass is 
fixed at the background level. 
However, in the presence of small perturbations of the scalar 
field the Planck and active gravitational mass are 
perturbed while inertial masses are not. This difference is 
critical for certain applications.
For example, it guarantees that CDM perturbations 
can be ascribed in the present framework to scalar field 
perturbations (essentially to $\lesssim 10^{-3}-10^{-4}$ 
fractional variations of $G$) 
on galactic and galaxy cluster scales, [1], [34]. 
Otherwise they could be gauged away and DM phenomena cannot be 
explained this way.

It can be readily verified that the Euler-Lagrange equation for 
the scale factor $a(\eta)$ that extremizes the action 
$\mathcal{I}_{FRW}$ is indeed the Friedmann equation 
$\mathcal{H}^{2}+K=a^{2}\rho_{M}
+\frac{\Lambda a^{2}}{3}+const./a^{2}$, where 
$\mathcal{H}\equiv a'/a$ is the conformal Hubble function, 
and it is assumed that 
$\mathcal{L}_{M}(a)$ is a power-law in $a$, much like $\rho_{M}$ 
is a power-law in standard cosmology. 
Here, $\mathcal{L}_{M}(a)$ accounts for all sources (e.g. dust, 
radiation, etc.) other than spatial curvature and cosmological 
constant (which are characterized by effective equations 
of state -1/3 and -1, respectively). 
Completing the derivation requires that the integration constant 
is related to the energy density of 
radiation $\rho_{r}=const./a^{4}$. 
Since $\rho_{M}>0$ it is clear that the 
kinetic and potential terms in equation (20) 
have the `wrong' relative sign 
(unless $K>0$ or $\Lambda<0$), yet the GR-based 
FRW model is not considered to be `plagued with ghosts', 
or `disastrous instabilities' at the classical level. 
On the contrary, FRW is the backbone of the remarkably 
successful {\it standard} cosmological model.

Re-defining the scale factor $\phi\equiv a$, 
the FRW action is reformulated as a WIST theory 
with a non-positive kinetic term, defined on a static background 
\begin{eqnarray}
\mathcal{I}_{FRW}&=&\int\left(\frac{1}{6}R\phi^{2}-\phi'^{2}
-\lambda\phi^{4}+\tilde{\mathcal{L}}_{M}(\phi)\right)\sqrt{-\gamma}d^{4}x.
\end{eqnarray}
Here, $\tilde{\mathcal{L}}_{M}=a^{4}\mathcal{L}_{M}$, $R=6K$ 
and $\lambda\equiv\Lambda/3$. 
The single degree of freedom of standard FRW spacetime, 
$a(\eta)$, is here replaced by the scalar field 
$\phi(\eta)$. Clearly, renaming the scale factor, 
$a\rightarrow\phi$, 
in going from the general relativistic equation (20) 
to the WIST form, equation (21), does not introduce 
any new dynamics, in particular no unacceptable new instabilities that 
are not already present in the FRW spacetime arise due to 
this field redefinition.
The relative `wrong' sign between the kinetic and potential 
term is simply manifested in that the integration 
of the Friedmann equation 
results in $a(\eta)$ [or equivalently $\phi(\eta)$] 
that evolves monotonically with $\eta$ rather than 
having oscillatory behavior.

As equation (20) and equation (21) 
are equivalent any perturbation is either 
$a(\eta)$ of equation (20) or $\phi(\eta)$ of 
equation (21) will have 
precisely the same dynamics. Exactly as in the standard 
cosmological model $\delta a$ is never considered as a dynamical 
degree of freedom, so should be the case with $\delta\phi$. 
This is so since $g_{\mu\nu}=a^{2}\gamma_{\mu\nu}$, 
and therefore any perturbation in $a$ can be absorbed in 
$\delta\gamma_{\mu\nu}$, a perturbation of $\gamma_{\mu\nu}$.
Considering scalar metric perturbations only -- these 
are the Newtonian potentials. Therefore, invoking WI implies that any 
perturbation in $G$ induces scalar metric perturbations [2], 
and vice versa any excessive gravitational 
potential could be explained by corresponding perturbations in $G$ [1].

As we just saw, local {\it isotropic} scaling has sufficient freedom 
to allow for reformulation of the (background) FRW model on 
a static background. 
However, this is not generally the case with Bianchi Universes. 
The latter are homogeneous but not isotropic. 
As such, they may have up to three 
different expansion rates along the three principal axes for which 
a corresponding number of functional degrees of freedom are required. For that 
purpose, the more general {\it anisotropic} local rescaling freedom described 
in section 2 is required. In addition, at the perturbation level, 
vector and tensor perturbation modes cannot be accounted for by a single 
function $\phi$, and for that purpose the more general theory described in 
section 2 is required as well, as was already mentioned above.

\section{Global Symmetry}

As reviewed in section 3 one of the often-made claims about WIST is that 
the procedure of `conformalizing' GR does not really result 
in modified physics. 
Specifically, it is argued, the conformal frame can always be so chosen so as 
to recover GR from equation (11) by setting $\phi$ to a constant. 
Irrespective of the validity of this claim in light of 
the discussion towards the end of section 2, it is clear 
that if a global symmetry is associated with the tensor field 
$\Phi_{\mu}^{\nu}$ then the presence of a global conserved charge 
definitely modifies the theory. Specifically, 
the field $\Phi_{\mu}^{\ \nu}$ itself 
can be generalized to carry the indices of an internal 
(non-coordinate) symmetry, i.e. $\Phi_{\mu}^{\ \nu}
\rightarrow{\bf \Phi}_{\mu}^{\ \nu}=(\Phi_{I}^{J})_{\mu}^{\ \nu}$ 
where boldface letters indicate that the tensor field carries internal indices 
(capital Latin letters) of a 
(typically unitary) group. The metric on (internal) group space 
is denoted $\mathcal{G}_{IJ}$, such that 
e.g., ${\bf \Phi}_{\mu}^{\ \alpha}{\bf \Phi}_{\alpha}^{\ \nu}$ 
is a shorthand notation for $(\Phi_{I}^{K})_{\mu}^{\ \alpha}\mathcal{G}_{KL}
(\Phi_{M}^{L})_{\alpha}^{\ \nu}$, and it is understood that contracted 
capital Latin letters are summed over. Equation (2) is then generalized to 
\begin{eqnarray}
g_{\mu\nu}&&\rightarrow\bar{g}_{\mu\nu}\equiv {\bf \Phi}_{\mu}^{\ \alpha}
g_{\alpha\beta}{\bf \Phi}_{\ \nu}^{\beta},\nonumber\\
\mathcal{L}_{M}&&\rightarrow\bar{\mathcal{L}}_{M}/|{\bf \Phi}|.
\end{eqnarray}
When expanded, the first of equations (22) is 
$g_{\mu\nu}\rightarrow\bar{g}_{\mu\nu}\equiv (\Phi_{I}^{K})_{\mu}^{\ \alpha}
g_{\alpha\beta}\mathcal{G}_{KL}(\Phi^{LI})_{\ \nu}^{\beta}$.
In the special case that the internal symmetry is described by a unitary 
group (whose elements are $U_{IJ}$)
then no stretch and contraction of the fundamental units takes place 
($|U|=1$), i.e. no local scale invariance, but rather only a global 
charge is added to the theory. In this case, derivatives 
of the new metric field generically result in quadratic kinetic terms 
of phases that function as `cyclic coordinates' on the group manifold, 
thereby resulting in conserved charges.

We stress that the internal symmetry 
is assumed to apply only to the field ${\bf \Phi}_{\mu}^{\ \nu}$, 
and not to any other fields. 
There is a conserved global charge that is associated with this global 
symmetry that physically alters the field equations of gravitation. 
The presence of this charge modifies the kinetic term associated with 
the field, as is demonstrated in e.g., the U(2) case in Appendix B.
An example of this within a WIST-based bouncing cosmological model 
is described in [34].

\section{Summary}
Symmetry plays a central role in our current understanding of 
the fundamental interactions. The SM of particle 
physics is based on a local SU(3)$\times$SU(2)$\times$U(1) 
gauge symmetry, and GR is diffeomorphism-invariant. 
In the present work, the idea that GR accommodates a broader symmetry, 
{\it anisotropic} local scale invariance (of which WI is just a special case), 
is explored first. 
The underlying idea is that not only that the gravitational 
interaction (and by induction also all the other interactions which 
take place in the spacetime arena) does not depend on the reference 
frame of the observer (as is manifested by diffeomorphism invariance of, 
e.g., the Einstein and geodesic equations) it also does not depend 
on ({\it a priori} arbitrary) choices of units, i.e. yardsticks, e.g. the 
Planck length scale, $l_{P}$. 
The latter is the natural yardstick in vacuum. In the non-vacuum case 
the natural meter stick would be the dynamical gravitational time scale, 
$(G\rho_{M})^{-1/2}\propto l_{P}^{-1}\rho_{M}^{-1/2}$ where $G$ is Newton 
constant and $\rho_{M}$ is the energy density of matter. 
The standard units choice 
adopted in GR is that both $G$ and particle masses (active gravitational 
masses included) are fixed constants. 

Newton constant, $G$, has been 
measured in our solar system at a reasonable precision, and is purportedly 
also deduced (at the few percent precision) outside of our 
solar system from Big Bang nucleosynthesis (BBN) via observational 
inference of light element abundances , etc. 
However, all these estimates of $G$ are based on 
the premise that all other universal constants are truly constants, as is 
clearly manifested in the Brans-Dicke program. In addition, Newtonian gravity 
fails to account for a few observed phenomena on galactic scales and 
beyond in spite of the fact that the gravitational pull is 
weak in these systems and Newtonian gravity is expected, 
at least naively, to provide a reasonably accurate description. 
The standard solution to this problem involves an assumption 
about the existence of a non-luminous DM substance, 
presumably exotic electrically neutral particles. 
Another way, albeit less popular, to tackle the 
problem has involved modifying the Newtonian 
dynamics on sufficiently large scales. Both remedies 
assume that $G$ is a fixed universal constant 
across space and time; in the absence of compelling 
direct evidence for the universality of $G$ this 
has to be assumed. 

However, positing WI as an additional symmetry of (what is 
usually considered GR) opens up new possibilities. 
Promoting $G$ to a tensor (section 2) or scalar 
(section 3) scale-deformation field necessitates that 
active gravitational masses are regulated by the same 
fields as well. Clearly, it is assumed here 
that (anisotropic or isotropic) local scale invariance 
applies only within the domain of validity 
of GR. This is of course not the case on microscopic scales. While GR 
is assumed here to be locally scale-invariant, the SM of particle physics 
is assumed to be as is, i.e. non-scale-invariant, 
with no modifications. The effective 
energy and pressure contained in spatio-temporal 
variations of the scalar field, i.e. $G$ and active gravitational 
masses, then provide additional (non-particulate) source for 
spacetime curvature, thereby accounting for, e.g. `DM phenomena' 
with no recourse to DM. In the present work we explored a few 
key aspects of the underlying locally scale-invariant 
generalizations of GR. In addition to local scale invariance, we 
equip the `scaling' fields ($\phi$ or $\Phi_{\mu\nu}$) 
with internal symmetry and generally non-vanishing global charge.  

It should be emphasized that whereas the case isotropic local 
rescaling has long been explored, its generalization, 
i.e., the {\it anisotropic} case that is discussed in 
section 2, has not been considered in the literature in 
full generality (except for the special case of Minkowski 
background where it is known as the `vierbein formalism') 
to the best of our knowledge.

While GR is modified in this framework, the SM is not. We stress that 
unlike in other frameworks that `conformalize' all the four 
fundamental interactions and so any `new physics' can be entirely 
gauged away by means of an appropriate Weyl transformation (in the 
absence of new global charges), this cannot 
be done in the framework adopted in the present work. It should be also 
emphasized that even if all the four fundamental interactions are 
`conformalized' the presence of internal symmetry and global charge 
do modify the field equations, i.e. the dynamics, and consequently 
give rise to `new physics'.

Whereas GR is recovered from the locally anisotropic scale-invariant 
or WIST theory in a given conformal frame 
(in case that $G$ is set to a constant), 
the additional functional degrees of freedom 
(obtained by promoting $m_{p}$ to a scalar $\phi$ or tensor $\Phi_{\mu\nu}$ 
field) provides an ample 
freedom that allows doing away with clustered DM as mentioned above. 
A notable property of $\phi$ is that it is formally a `ghost' field, 
i.e. its kinetic and potential terms have the `wrong' relative sign. 
However, this property by itself does not disqualify the theory 
because the problem with ghost-afflicted theories is 
that the field configuration {\it dynamically} runs away from its 
ground state, or equilibrium, by ever lowering the kinetic energy 
while increasing the potential term. However, as was explicitly 
shown in section 3, and as is equally clear from the 
underlying WI, the spacetime-dependent $\phi$ in the case of 
WIST, or $\Phi_{\mu}^{\nu}$ in the anisotropic case, is non-dynamical; 
it is simply not required to obey any specific Euler-Lagrange 
equation. Consequently, 
physical systems described by locally scale-invariant theories of 
gravity are not doomed to runaway from stable 
field configurations. We also illustrated this conclusion with 
the redshifting Universe in section 4. The FRW spacetime can be recast in the canonical 
WIST form on a static background. Cosmological redshift is then explained not 
by space expansion but rather by contraction of our fundamental yardsticks, 
e.g. the Rydberg `constant', which are (directly and indirectly) 
regulated by $\phi$. 
Whereas $\phi$ can be locally rescaled along with a corresponding 
rescaling of $g_{\mu\nu}$, adopting a specific conformal frame in 
which the metric is static implies that $\phi$ satisfies exactly 
the same Friedmann equation which is normally satisfied by $a$, 
the scale factor that describes space expansion in the standard 
cosmological model. 
The conformal Hubble function $\mathcal{H}=a'/a$ is now replaced 
by $\phi'/\phi$. In the WIST version of the FRW action 
the negative kinetic term has been actually increasing ($\phi'/\phi$ 
monotonically decreases) asymptotically towards zero over the entire 
cosmic history between the `Big Bang' and the present time. 
This is a counter-example to often-raised concerns 
about instabilities associated with WIST. It is true that on the 
face of it we could have found ourselves in a contracting Universe 
with an ever decreasing negative kinetic term of either $a$ or $\phi$ 
(in the GR and WIST formulations respectively) culminating in a 
`big crunch', but even in this case the WIST dynamics is not `worse' 
than that of GR. More general cosmological models, e.g. Bianchi-types 
models, that are described by non-isotropic expansion rates are in general 
described by three scale factors. This family of models can be easily 
described by the {\it anisotropic} locally scale-invariant theory.

\section*{Appendix A: The Weak gravitational field limit as an inertial force}

The classical trajectories followed by particles in GR are described 
by geodesics
\begin{eqnarray}
\ddot{x}^{\mu}+\Gamma^{\mu}_{\rho\sigma}\dot{x}^{\rho}\dot{x}^{\sigma}=0, 
\end{eqnarray}
where $\Gamma^{\mu}_{\rho\sigma}$ is the Christoffel symbol that is 
calculated from the metric field and its derivatives in the usual way, 
and $\dot{x}^{\mu}\equiv \frac{dx^{\mu}}{d\lambda}$, with $\lambda$ 
an affine parameter. Equation (23) is subject to the constraint 
$g_{\mu\nu}\dot{x}^{\mu}\dot{x}^{\nu}=c$ where $c$ equals -1 or 0 
in case of massive or massless particles, respectively.
In the presence of additional non-gravitational forces, $F^{\mu}$, 
equation (23) generalizes to
\begin{eqnarray}
\ddot{x}^{\mu}+\Gamma^{\mu}_{\rho\sigma}\dot{x}^{\rho}\dot{x}^{\sigma}
=\frac{F^{\mu}}{m}, 
\end{eqnarray}
in the case of a massive particle of mass $m$.
Now, the Christoffel symbol transforms inhomogeneously 
under Weyl transformations (equation 12)
\begin{eqnarray}
\Gamma^{\mu}_{\rho\sigma}\rightarrow
\tilde{\Gamma}^{\mu}_{\rho\sigma}=\Gamma^{\mu}_{\rho\sigma}
+\Omega^{-1}(\delta_{\rho}^{\mu}\Omega_{\sigma}
+\delta_{\sigma}^{\mu}\Omega_{\rho}-g_{\rho\sigma}g^{\mu\delta}\Omega_{\delta}).
\end{eqnarray}
Therefore, in the case $F^{\mu}=0$, equation (23) is rewritten in the 
new conformal frame as
\begin{eqnarray}
\ddot{x}^{\mu}+\tilde{\Gamma}^{\mu}_{\rho\sigma}
\dot{x}^{\rho}\dot{x}^{\sigma}=\frac{2\Omega_{\sigma}}{\Omega}
\dot{x}^{\sigma}\dot{x}^{\mu}-cg^{\mu\nu}\Omega_{\nu}/\Omega,
\end{eqnarray}
or equivalently
\begin{eqnarray}
\ddot{x}^{\mu}+\tilde{\Gamma}^{\mu}_{\rho\sigma}
\dot{x}^{\rho}\dot{x}^{\sigma}=-\frac{2\phi_{\sigma}}{\phi}
\dot{x}^{\sigma}\dot{x}^{\mu}+cg^{\mu\nu}\phi_{\nu}/\phi. 
\end{eqnarray}
We see that what is considered as a pure gravitational force 
in one conformal frame is interpreted as not pure gravitational 
in other frames. A striking example is the case where 
spacetime is described by Minkowski metric in the new frame, 
and in cartesian coordinates then 
$\tilde{\Gamma}^{\mu}_{\rho\sigma}$ vanishes. 
In this particular case equation (27) describes a purely non-gravitational 
interaction. The particles (massive or massless) follow 
exactly the same geodesics (that describe e.g. gravitational clustering 
of massive particles or light bending of massless photons) in the 
two conformal frames, but while in the first frame it is due to spacetime 
curvature in the other frame it is due to an inertial `force', 
$F\equiv -2\dot{x}^{\mu}\dot{x}^{\nu}\phi_{\nu}/\phi
+c\eta^{\mu\nu}\phi_{\nu}/\phi$, 
where $\eta^{\mu\nu}$ is Minkowski metric.

For example, the line element that describes a 
weakly perturbed Minkowski spacetime  
in Cartesian coordinates and in case that stress is 
negligible, and neglecting vector and tensor modes, 
is $ds^{2}=-(1+2\varphi)dt^{2}+(1-2\varphi)dx^{i}dx^{j}$.
Here $\varphi(x)$ is the gravitational potential.
Applying a Weyl transformation $\Omega=1+\varphi$ 
the metric in the new conformal frame becomes purely Minkowski, 
and $\tilde{\Gamma}^{\mu}_{\rho\sigma}$ vanishes in Cartesian 
coordinates. However, the r.h.s of equation (27) does not vanish now. 
The entire kinematics of test particles is now due to the fact that 
$\phi_{0}\rightarrow\tilde{\phi}\approx\phi_{0}[1-\varphi(x)]$. 
In other words, light bending and gravitational clustering are 
not due to spacetime curvature (which is flat in this case) 
but rather due to the spacetime dependence of $G$ and 
active gravitational masses.
In the presence of vector and/or tensor modes the single functional degree 
of freedom, $\phi$, added by the WIST theory is insufficient for entirely 
flattening the spacetime geometry and the more general locally 
scale invariant theory that is described in section 2 is required. 
The tensor field $\Phi_{\alpha\beta}$ can be used reformulate GR on a Minkowski 
spacetime, in which case it reduces to the vierbein field. The generalization of 
equation (27) in that case is straightforward.

\section*{Appendix B: The Case of U(2) Global Symmetry}

As an example for possibilities opened up by introducing 
internal symmetries associated 
with the gravitational interaction, we consider the case that the scalar 
field $\boldsymbol{\phi}$ is a U(2) multiplet. Elements of this group 
are represented by
\begin{equation}
  \boldsymbol{\phi} =\rho
  \left( {\begin{array}{cc}
    \cos(\theta)e^{i\varphi} & \sin(\theta)e^{i\chi} \\
    -\sin(\theta)e^{-i\chi} & \cos(\theta)e^{-i\varphi} \\
  \end{array} } \right),
\end{equation}
where $\rho$ is the modulus of the field and $\theta$, 
$\varphi$ and $\chi$ are three phases.
The line element on this group manifold is 
$ds_{\boldsymbol{\phi}}^{2}=\mathcal{G}_{IJ}d\phi^{I}d\phi^{J}$, where 
contracted capital Latin indices are summed over. Since this is a unitary group 
then, 
$ds_{\boldsymbol{\phi}}^{2}=\frac{1}{2}Tr(d\boldsymbol{\phi}^{\dagger}
d\boldsymbol{\phi})=
-\frac{1}{2}Tr(\boldsymbol{\phi}^{-1}d\boldsymbol{\phi})^{2}
=d\rho^{2}+\rho^{2}\left[\cos^{2}(\theta)d\varphi^{2}
+\sin^{2}(\theta)d\chi^{2}+d\theta^{2}\right]$. 
The lagrangian of the kinetic term, 
$\mathcal{L}_{kin}=\mathcal{G}_{IJ}\phi_{\mu}^{I}\phi^{\mu,J}$ 
(which is a straightforward generalization of the 
kinetic term in Equation 11) is therefore 
$\mathcal{L}_{kin}=-\dot{\rho}^{2}-\rho^{2}[\cos^{2}(\theta)\dot{\phi}^{2}
+\sin^{2}(\theta)\dot{\chi}^{2}+\dot{\theta}^{2}]$. 
The Euler-Lagrange equations associated with $\chi$ and $\varphi$ 
that are obtained from variations of $\mathcal{L}_{kin}$ 
can be each integrated once to give
\begin{eqnarray}
\dot{\chi}&=&\frac{c_{\chi}}{\rho^{2}\sin^{2}\theta}\nonumber\\
\dot{\varphi}&=&\frac{c_{\varphi}}{\rho^{2}\cos^{2}\theta},
\end{eqnarray}
where $c_{\chi}$ and $c_{\varphi}$ are integration constants. 
Employing these in the field equation for $\theta$ 
and integrating once we obtain
\begin{eqnarray}
\rho^{4}\dot{\theta}^{2}=-\frac{c_{\chi}^{2}}{\sin^{2}\theta}
-\frac{c_{\varphi}^{2}}{\cos^{2}\theta}+c_{\theta}^{2}, 
\end{eqnarray}
where $c_{\theta}$ is another integration constant.
Employing all these in the kinetic 
term $\phi_{\mu}\phi^{\mu}=-\dot{\rho}^{2}
-\rho^{2}[\cos^{2}(\theta)\dot{\varphi}^{2}+\sin^{2}(\theta)\dot{\chi}^{2}
+\dot{\theta}^{2}]$ we obtain 
$-\dot{\rho}^{2}-\frac{c_{\theta}^{2}}{\rho^{2}}$, and both integration 
constants $c_{\chi}$ and $c_{\varphi}$ vanish 
from our final expression. All this amounts to 
replacing $\phi$ with $\rho$ everywhere in equation (11) 
in addition to introducing a negative contribution 
to the potential, $-\frac{c_{\theta}^{2}}{\rho^{2}}$. 

In the cosmological context this new term 
competes with the radiation term at sufficiently small $\rho$. 
This term, obtained due to the U(2) symmetry of the scalar field 
(a symmetry which is not recognized by GR) rather than having to 
recourse to exotic forms of matter with negative energy density, 
is responsible to a bounce, where the cosmological expansion rate 
momentarily halts. As expected, this is exactly the 
result obtained in [34] in case of U(1) 
symmetry. The constant $c_{\theta}$ is simply a global charge associated 
with the U(1) subgroup, which is the relevant symmetry in a `central' potential, 
i.e. a potential which is independent on the phase, and depends only on $\rho$. 
In cosmology the term $\propto c_{\theta}$ serves as an effective 
stiff matter with a negative energy density, thereby replacing the initial 
Big Bang singularity with a Big Bounce [34].

\section*{References}
\begin{enumerate}[label={[\arabic*]}]
\item Shimon, M.: Weyl-Invariant Gravity and the Nature of Dark Matter. 
Class. Quantum Gravity {\bf 38} 085001 (2021)
\item Shimon, M.: Possible Resolution of the Hubble Tension with 
Weyl Invariant Gravity. JCAP, {\bf 2022}, 048 (2022)
\item Weinberg, S.: The cosmological constant problem. 
Reviews of Modern Physics {\bf 61} 1 (1988)
\item Deser, S.: Scale invariance and gravitational 
coupling. Ann. Phys. {\bf 59} 248 (1970)
\item Anderson, J.~L.: Scale Invariance of the Second Kind 
and the Brans-Dicke Scalar-Tensor Theory. Phys. Rev. D {\bf 3} 1689 (1971) 
\item Freund, P.~G.~O.: Local scale invarlance and gravitation. 
Annals of Physics {\bf 84} 440 (1974)
\item Kallosh, R.: On the renormalization problem of 
quantum gravity. Physics Letters B {\bf 55} 321 (1975)
\item Englert, F., Gunzig, E., Truffin, C., et al.: 
Conformal invariant general relativity with dynamical 
symmetry breakdown. Phys. Lett. B {\bf 57} 73 (1975)
\item Smolin, L.: Gravitational radiative corrections as 
the origin of spontaneous symmetrybreaking! Phys. Lett.B {\bf 93} 95 (1980)
\item Padmanabhan, T.: LETTER TO THE EDITOR: 
Conformal invariance, gravity and massive gauge theories. 
Class. Quantum Gravity {\bf 2} L105 (1985)
\item Edery, A., Fabbri, L. \& Paranjape, M. B.: 
Spontaneous breaking of conformal invariance intheories 
of conformally coupled matter and Weyl gravity. Class. Quantum 
Gravity {\bf 23} 6409 (2006) [hep-th/0603131]
\item 't Hooft, G.: A Class of Elementary Particle Models 
Without Any Adjustable Real Parameters. 
Foundations of Physics {\bf 41} 1829 (2011) [arXiv:1104.4543]
\item Bars, I., Chen, S.-H., Steinhardt, P.~J., 
et al.: Complete set of homogeneous isotropic analytic 
solutions in scalar-tensor cosmology with radiation 
and curvature. Phys. Rev. D {\bf 86} 083542 (2012) [arXiv:1207.1940]
\item Bars, I., Chen, S.-H., Steinhardt, P.~J., et al.: 
Antigravity and the big crunch/big bang transition, 
Physics Letters B {\bf 715} 278 (2012) [arXiv:1112.2470]
\item Kallosh, R., Linde, A.: Universality class 
in conformal inflation. J. Cosmol. Astropart. Phys. {\bf 2013} 
002 (2013) [arXiv:1306.5220]
\item Bergmann, P.~G.: Comments on the scalar-tensor theory. 
Int. J. Theor. Phys. {\bf 1} 25 (1968)
\item Wagoner, R.~V.: Scalar-Tensor Theory and 
Gravitational Waves. Phys. Rev. D {\bf 1} 3209 (1970)
\item Chamseddine, A.~H. \& Mukhanov, V.: Mimetic Dark Matter. 
Jour. of High Energy Phys. {\bf 2013} 135 (2013) [arXiv:1308.5410]
\item Brans, C., Dicke, R.~H.: Mach's Principle and a Relativistic 
Theory of Gravitation. Physical Review {\bf 124} 925 (1961)
\item Bertotti, B., Iess, L., Tortora, P.: A test of general relativity using radio links with the Cassini spacecraft, 
Nature {\bf 425} 374 (2003)
\item Bars, I., Steinhardt, P., Turok, N.: Local conformal symmetry in physics and cosmology, Phys. Rev. D {\bf 89} 043515 (2014) [arXiv:1307.1848]
\item Cline, J.~M., Jeon, S., \& Moore, G.~D.: The phantom menaced: Constraints on low-energy effective ghosts. Phys. Rev. D {\bf 70} 043543 (2004) [hep-ph/0311312]
\item Faraoni, V., Nadeau, S.: (Pseudo)issue of the conformal frame 
revisited. Phys. Rev. D {\bf 75} 023501 (2007) [gr-qc/0612075]
\item Flanagan, {\'E}. {\'E}: REPLY TO COMMENT: The conformal frame freedom in theories of gravitation. Class. Quantum Gravity {\bf 21} 3817 (2004) [gr-qc/0403063]
\item Jackiw, R., Pi, S.-Y.: Fake conformal symmetry in conformal cosmological models. Phys. Rev. D {\bf 91} 067501 (2015) [arXiv:1407.8545]
\item Tsamis, N.~C., Woodard, R.~P.: No new physics in conformal scalar-metric theory. Annals of Physics {\bf 168} 457 (1986)
\item Ohanian H.~C.: Weyl gauge-vector and complex dilaton 
scalar for conformal symmetry and its breaking. 
Gen. Rel. Grav. {\bf 48} 25 (2016) [arXiv:1502.00020]
\item Carlip, S.: Is quantum gravity necessary? 
Classical and Quantum Gravity {\bf 25} 154010 (2008) [arXiv:0803.3456]
\item Barrow J.~D., Kimberly D., Magueijo J.: Bouncing universes with 
varying constants. Class. Quantum Gravity {\bf 21}, 4289 (2004)
\item Bondi, H.: Negative Mass in General Relativity, 
Rev. Mod. Phys. {\bf 29} 423 (1957)
\item Avelino, P.~P., Azevedo, R.~P.~L.: Perfect fluid Lagrangian and 
its cosmological implications in theories of gravity with nonminimally coupled 
matter fields. Phys. Rev. D {\bf 97} 064018 (2018) [arXiv:1802.04760]
\item Dicke, R.~H.: Mach's Principle and Invariance under 
Transformation of Units. Physical Review {\bf 125} 2163 (1962)
\item Noerdlinger, P.~D.: Theoretical Necessity of a Positive 
Value for the Brans-Dicke Coupling Constant $\omega$, {\it Physical Review} {\bf 170} 1175 (1968)
\item Shimon, M.: arXiv:2205.07251 (2022)


\end{enumerate}

\end{document}